\documentclass[%
 reprint,
 amsmath,amssymb,
 aps,
]{revtex4-2}

\usepackage{tikz}
\usepackage{flushend}
\usepackage{verbatim}
\usepackage{verbatim,upgreek}
\usepackage{chngcntr}
\usepackage{graphicx}
\usepackage{dcolumn}
\usepackage{bm}
\usepackage{comment}
\usepackage{amsmath}
\usepackage{hyperref}
\hypersetup{
    colorlinks=true,
    citecolor=blue,
    }

\usepackage{color}
\usepackage{physics}

\usepackage{lineno}

\begin{document}

\newcommand{\ms}[1]{\mbox{\scriptsize #1}}
\newcommand{\msb}[1]{\mbox{\scriptsize $\mathbf{#1}$}}
\newcommand{\msi}[1]{\mbox{\scriptsize\textit{#1}}}
\newcommand{\nn}{\nonumber} 
\newcommand{\dg}{^\dagger}

\preprint{APS/123-QED}

\title{A spin-refrigerated cavity quantum electrodynamic sensor}

\author{Hanfeng Wang$^{1}$}

\author{Kunal L. Tiwari$^{2}$}
 
\author{Kurt Jacobs$^{3,4}$}

\author{Michael Judy$^{5}$}

\author{Xin Zhang$^{5}$}

\author{Dirk R. Englund$^{1,6}$}
 \email{englund@mit.edu}
 
\author{Matthew E. Trusheim$^{1,3,6}$}
 \email{mtrush@mit.edu, matthew.e.trusheim.civ@army.mil}
 
\affiliation{%
$^{1}$ Research Laboratory of Electronics, M.I.T, 50 Vassar Street, Cambridge, MA 02139, USA\\
$^{2}$ MIT Lincoln Laboratory, Lexington, MA 02421, USA\\
$^{3}$ DEVCOM Army Research Laboratory, Adelphi, MD 20783, USA\\
$^{4}$ Department of Physics, University of Massachusetts Boston, MA 02125, USA\\
$^{5}$ Analog Devices, Inc., 1 Analog Way, Wilmington, MA 01887, USA \\
$^{6}$ MIT Institute of Soldier Nanotechnology, 500 Technology Square, Cambridge, MA 02139, USA }

\date{\today}

\begin{abstract}

Quantum sensors based on solid-state defects, in particular nitrogen-vacancy (NV) centers in diamond, enable precise measurement of magnetic fields, temperature, rotation, and electric fields.
However, the sensitivity of leading NV spin ensemble sensors remains far from the intrinsic spin-projection noise limit.
Here we move towards this quantum limit of performance by introducing (i) a cavity quantum electrodynamic (cQED) hybrid system operating in the strong coupling regime, which enables high readout fidelity of an NV ensemble using microwave homodyne detection;
(ii) a comprehensive nonlinear model of the cQED sensor operation, including NV ensemble inhomogeneity and optical polarization; and
(iii) ``spin refrigeration'' where the optically-polarized spin ensemble sharply reduces the ambient-temperature microwave thermal noise, resulting in enhanced sensitivity. 
Applying these advances to magnetometry, we demonstrate a broadband sensitivity of 580 fT/$\sqrt{\mathrm{Hz}}$ around 15 kHz in ambient conditions.
We then discuss the implications of this model for design of future magnetometers, including devices approaching 12 fT/$\sqrt{\mathrm{Hz}}$ sensitivity.
Applications of these techniques extend to the fields of gyroscope and clock technologies.
\end{abstract}

\maketitle

\section{\label{sec:1}Introduction\protect}

Quantum sensors offer the prospect of device operation at the physical limit of performance \cite{degen2017quantum}. 
Among the many quantum sensing systems, nitrogen-vacancy (NV) centers in diamond have emerged as a promising platform \cite{barry2020sensitivity, clevenson2015broadband,doherty2013nitrogen,rovny2022nanoscale} due to favorable attributes including room-temperature spin polarization and readout, atomic-scale size~\cite{du2017control,li2023nanoscale,wang2022noninvasive,pelliccione2016scanned}, and long coherence times~\cite{bauch2018ultralong, stanwix2010coherence}.
NV-based sensors use resonance spectroscopy of the ground-state spin triplet transition frequencies to infer environmental properties, yielding excellent performance in a variety of sensing modalities including magnetometry \cite{taylor2008high, fescenko2020diamond, wolf2015subpicotesla}, electrometry \cite{dolde2011electric,wang2020electrical,wang2023field} and inertial sensing \cite{jarmola2021demonstration,soshenko2021nuclear}, even in extreme regimes \cite{fu2020sensitive}. For spin ensemble magnetometers, the sensitivity can be written as $\eta =  \sigma_{\mathrm{e}}/\gamma_{\mathrm{e}}\sqrt{NT_2^*}$, where $\gamma_{\mathrm{e}}$ is the gyromagnetic ratio, $T_2^*$ is the spin dephasing time, $\sigma_{\mathrm{e}}$ is the inverse readout fidelity, and $N$ is the number of spins. The readout fidelity for NV-ensemble sensors using continuous optical approaches is typically a factor of $\sigma_{\mathrm{e}}\approx 5000$ above the quantum spin-projection limit \cite{barry2020sensitivity,barry2016optical,schloss2018simultaneous}, primarily due to limitations in fluorescence collection efficiency, and is limited to a factor of order $100$ even in the case of nearly ideal collection \cite{barry2023sensitive} due to imperfect optical selection rules.

\begin{figure*}[t!]
\includegraphics[width = 1\textwidth]{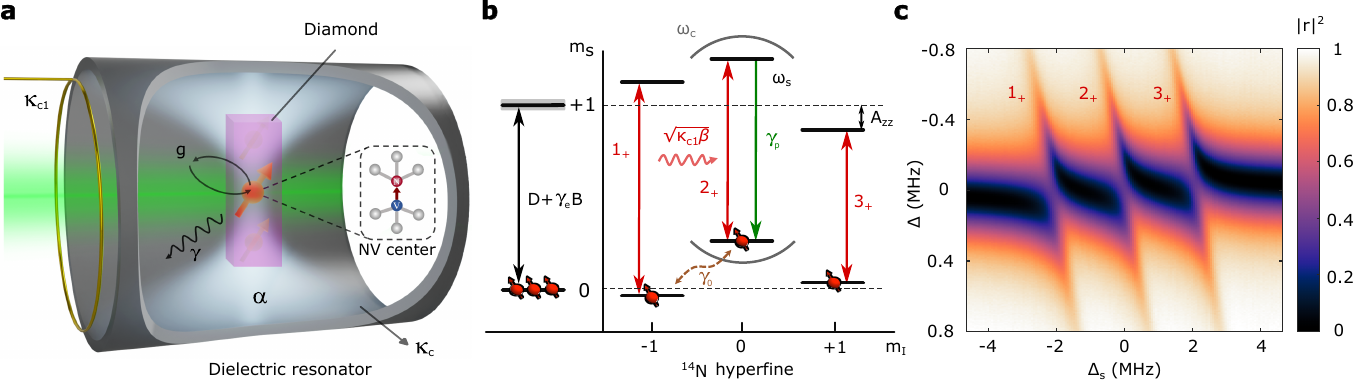}
\caption{\label{fig:1} \textbf{Hybrid NV-cavity system} \textbf{a}, A diamond containing an NV ensemble is located at the mode maximum of the TE01$\delta$ mode of a dielectric resonator, leading to a coupling strength of $g$.
A green laser is applied to continuously polarize the NV spins to the $|m_s=0\rangle$ electronic ground state, and a detection loop is incorporated to electrically probe the system.
The cavity possesses an intrinsic loss rate of $\kappa_{\ms{c}}$ and a coupling loss rate of $\kappa_{\ms{c}1}$, while the spin ensemble has an inhomogeneous linewidth $\Gamma$ and a homogeneous linewidth $\gamma$.
\textbf{b}, NV energy level structure.
The electronic spin has a splitting of $D+\gamma_e B$.
Three transitions $1_+$, $2_+$, and $3_+$ shown due to the hyperfine coupling with $^{14}N$.
\textbf{c}, Reflection $|r|^2$ with cavity detuning $\Delta = \omega_{\mathrm{d}}-\omega_{\mathrm{c}}$ and $\Delta_{\mathrm{s}} = \omega_{\mathrm{s}} -\omega_{\mathrm{c}}$.
Three strong coupling features are detected in the room temperature.}
\end{figure*}

To overcome the limitations of optical readout, the NV spin resonance can be probed in the microwave domain by coupling the NV ensemble to a microwave cavity mode~\cite{eisenach2021cavity}.
The sensor performance is then determined by the microwave spectroscopic response of the hybrid NV-cavity system, which can be precisely measured using established techniques. This approach eliminates photon shot noise as the dominant systematic noise, instead reaching the Johnson-Nyquist limit of the resonant microwave field. For sensors in ambient conditions, the thermal noise represents a fundamental limit set by the temperature of the apparatus, namely $k_\mathrm{B}T$. NV centers, however, have recently broken this barrier by serving as a mode-selective “spin refrigerator” that cools the detection channel below the Johnson-Nyquist noise \cite{fahey2023steady}, reaching towards the quantum spin-projection limit. In this work we show how this effect can be applied to enhance the performance of an NV ensemble magnetometer. 

Operating as a magnetometer in the strong coupling regime of cavity quantum electrodynamics (cQED) for the first time, our device achieves a sensitivity of 580 fT$/\sqrt{\mathrm{Hz}}$ corresponding to a continuous-wave inverse readout fidelity of $\sigma_{\mathrm{e}}\sim 400$. 
Our sensitivity is enhanced by spin refrigeration, reducing the effective voltage noise temperature by $\Delta T= 166$ K.  
Based on a comprehensive model describing the behavior of the cQED sensor (including inhomogeneity and nonlinear effects), we outline sensitivity regimes for varying spin ensemble properties, cavity designs, and power requirements.
We argue for the feasibility of 12 fT$/\sqrt{\mathrm{Hz}}$ sensitivity in an optimized device.
While the present work focuses on magnetometry, these results are broadly applicable across NV-enabled sensing modalities including inertial sensing and timekeeping \cite{soshenko2021nuclear,jarmola2021demonstration,trusheim2020polariton}.

\section{Principle of Device Operation}

\begin{figure*}
\includegraphics[width = 0.9\textwidth]{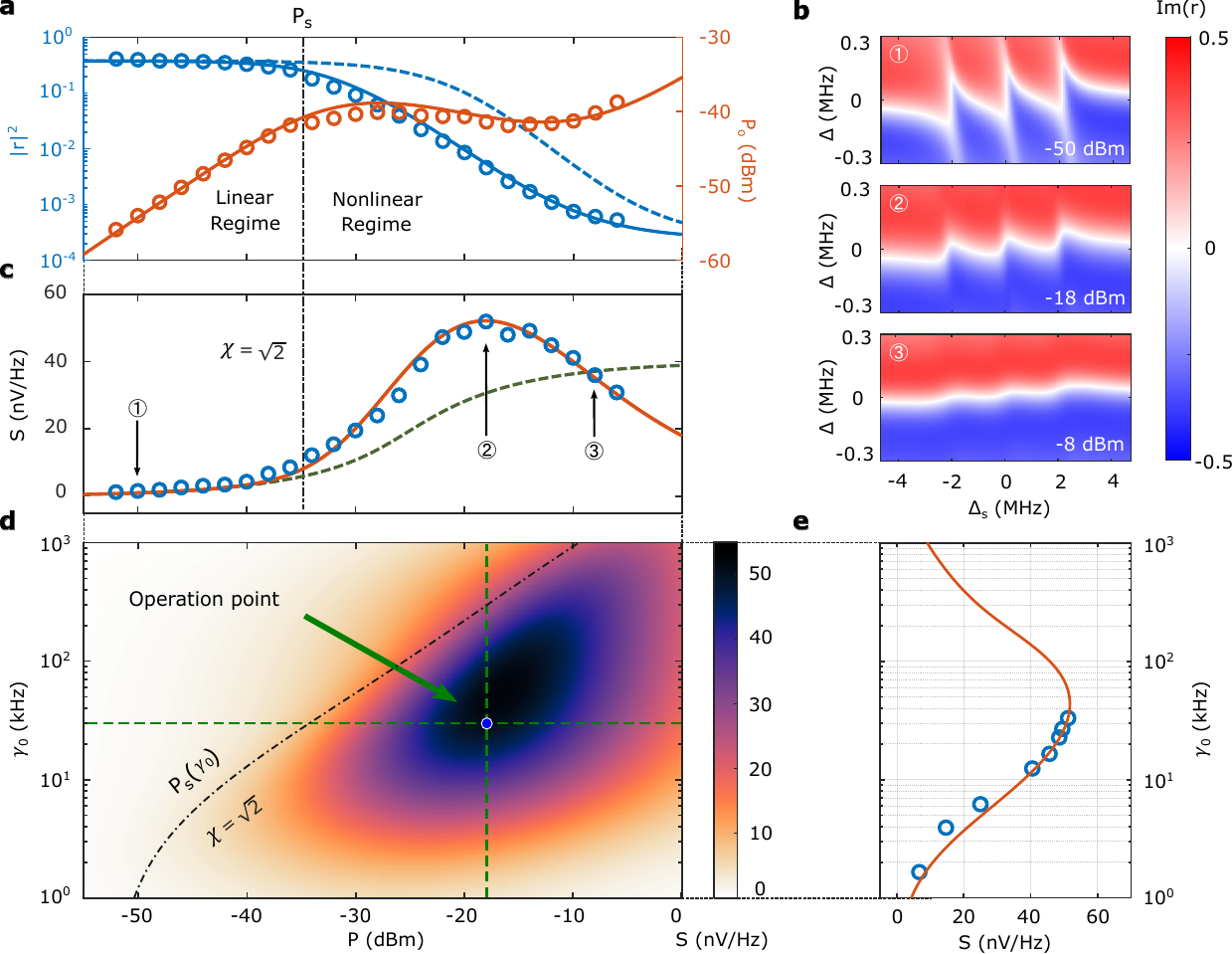}
\caption{\label{fig:2} \textbf{cQED sensor nonlinear model}.
\textbf{a}, Power reflection coefficient $|r|^2$ (blue) and reflected power $P_0$ (red) with resonant tuning ($\Delta=\Delta_{\mathrm{s}}=0$).
Circles are experimental data, solid lines are model results using best-fit parameters $\kappa_{\ms{c}1} = 2\pi\times125$ kHz, $\kappa_{\ms{c}} = 2\pi\times130$ kHz, $\Gamma = 2\pi\times 330$ kHz, $\gamma = 2\pi\times33$ kHz, $g$ = $2\pi\times190$ kHz.
The dashed line shows the model result assuming zero inhomogeneous broadening. 
\textbf{b}, Experimental imaginary component of the cavity reflection coefficient as a function of probe ($\Delta$) and atom-cavity ($\Delta_{\mathrm{s}}$) detunings for microwave input powers $P=-$50 dBm, $-18$ dBm, and $-8$ dBm.
\textbf{c}, Device signal $S$ as a function of input microwave power at fixed optical polarization rate $\gamma_0 = 2\pi\times30$ kHz (8 W of incident green laser power). Blue dots: experimental data. 
Red line: theoretical plot from nonlinear theory using the parameters given above. 
Green dotted line: the signal $S$ predicted by a thermally-polarized exceptional point model presented in previous work \cite{zhang2021exceptional}. The indicated points 1,2,3 correspond to the spectra shown in \textbf{b}.
\textbf{d}, Device signal $S$ as a function of input microwave power and optical polarization rate. The operating point that maximizes signal S is indicated with the blue dot
\textbf{e}, $S$ as a function of optical polarization rate at microwave power $P = -$18 dBm. The nonlinear model is the red curve and experimental measurements are blue points. }
\end{figure*}

\subsection{System Description}

We treat the hybrid system of an NV-doped diamond within a high-quality-factor ($Q$) dielectric resonator, shown schematically in Fig.~\ref{fig:1}a,b, with an inhomogeneous open Tavis-Cummings model.
Numerous prior studies have investigated similar systems in various parameter regimes \cite{kubo2010strong,bienfait2016reaching,schuster2010high,guo2023strong,zhang2014strongly,angerer2017ultralong,gimeno2020enhanced,wallraff2005approaching,loubser1978electron,tseitlin2010combining,frisk2019ultrastrong}. The $N\gg1$ NV centers are modeled as non-interacting two-level systems with transition frequencies $\omega_j$, distributed inhomogeneously due to heterogeneous local magnetic and strain environments as well as hyperfine coupling with $^{14}$N nuclear spins.
The spin relaxation rate $\gamma = \gamma_0 + \gamma_{\ms{p}}$ for uniform optical polarization rate $\gamma_{\ms{p}}$ (see Methods) and thermalization rate $\gamma_0\ll\gamma_{\ms{p}}$.
The cavity mode has loaded relaxation rate $\kappa = \kappa_{\ms{c}} + \kappa_{\ms{c}1}$ for intrinsic relaxation rate $\kappa_{\ms{c}}$ and coupling strength $\kappa_{\ms{c}1}$ to a microwave probe line.
Finally, we assume uniform coupling strength $g_{\mathrm{s}}$ between each individual spin and the cavity mode.

Under an input microwave drive field $\beta_{\mathrm{in}}$, the operator expectation values of the cavity field $\alpha$, spin coherence $s_j$, and excited-state population $p_j$ obey semi-classical equations of motion:~\cite{scully1999quantum, berman2011principles, gardiner1985inputoutput}
\begin{equation}
\begin{aligned}
\dot{\alpha} =&~-\qty(i\Delta + \frac{\kappa}{2})\alpha -ig_{\mathrm{s}}\sum_{j}s_j+\sqrt{\kappa_{\ms{c}1}}\beta_{\mathrm{in}}\\
\dot{s}_j =&~-\qty(i\Delta_j+\frac{\gamma}{2})s_j - ig_{\mathrm{s}}\qty(1-2p_j)\alpha\label{eq:sj_eom}\\
\dot{p}_j =&~-\gamma_\mathrm{p} p_j + ig_{\mathrm{s}}\qty(s_j\alpha^* - s_j^*\alpha).
\end{aligned}
\end{equation}
with $\Delta_j = \omega_\mathrm{d} - \omega_j$ the detuning between drive frequency $\omega_\mathrm{d}$ and NV transition frequency $\omega_j$, and $\Delta = \omega_\mathrm{d} - \omega_\mathrm{c}$ the detuning between drive frequency and cavity frequency $\omega_\mathrm{c}$.
The reflected signal $\beta_{\mathrm{out}}$ obeys the input-output relation $\beta_{\mathrm{out}} = \sqrt{\kappa_{\ms{c}1}}\alpha - \beta_{\mathrm{in}}$.
The response of $r=\beta_{\mathrm{out}}/\beta_{\mathrm{in}}$ to environmental changes which modify the spin transition frequencies allows the device to operate as a sensor.

To reach optimal measurement fidelity, a natural approach is to maximize microwave probe power and therefore signal to noise ratio.
In this regime, however, the strong microwave drive broadens the NV lines, reduces spin polarization, and the reflection coefficient acquires a dependence on probe power.
While the nonlinear spectroscopic properties of related homogeneous systems has been established~\cite{zhang2021exceptional}, the ensemble inhomogeneous distribution plays a central role in determining the reflection coefficient in this strong-driving regime.
Previous cavity-coupled sensor reports have noted the importance of ensemble inhomogeneity in determining performance, and have treated this problem phenomenologically~\cite{eisenach2021cavity, wilcox2022thermally}.
Related work describing bistability and critical dynamical behavior for transiently-driven inhomogeneous systems~\cite{angerer2017ultralong} are also well described by numerical treatments~\cite{krimer2019critical, zens2019bistablequantum}.
Here we extend this work in the context of sensing, determining the parametric dependence of the system response and saturation threshold on the inhomogeneous linewidth at the frequency tuning relevant to device operation.

\subsection{Linear Response and Strong Coupling}

In the limit of weak microwave drive such that all of the spins remain polarized ($p_j \ll 1$ for all $j$), the reflection coefficient is independent of drive amplitude and is given by the standard linear response expression:
\begin{align}
r = -1 + \frac{\kappa_{\ms{c}1}}{\kappa/2 + i\Delta + g^2\int\frac{P\qty(\omega')}{\gamma/2 + i(\omega-\omega')}\dd{\omega'}}
\label{eq:r_linear}
\end{align}
Here $g=g_{\mathrm{s}}\sqrt{N}$ is the strength of the collective coupling between the spin ensemble and the cavity, and $P\qty(\omega)$ is the normalized NV density per unit frequency with a center frequency of $\omega_\mathrm{s}$.

Generally, the exact form of the inhomogeneity may impact the shape of the spin-cavity polariton in a non-trivial manner, including potentially advantageous effects such as linewidth narrowing at very strong coupling~\cite{kurucz2011linearinhomogeneous, 2014_Putz}.
When spin, cavity, and drive are tuned to be nearly resonant ($\omega_{\mathrm{d}} = \omega_{\mathrm{c}}$, $\abs{\omega_{\mathrm{s}} - \omega_{\mathrm{c}} } \ll \gamma$), however, the reflection coefficient depends only on the magnitude and curvature of the inhomogeneous distribution at $\omega_\mathrm{s}$ (see Supplementary Material Sec. I). 
On the basis of this observation, we approximate the sub-ensemble inhomogeneous distribution as Lorentzian.
This allows us to explicitly evaluate the integral in Eq.~\eqref{eq:r_linear}, which can be expressed in terms of special functions for other inhomogeneous distributions (see Supplementary Material Sec. I).

Using this model, we parameterize our system from the experimental power reflection spectrum [Fig.~\ref{fig:1}c] (see Methods).
The three resolved anti-crossings correspond to the three $^{14}$N hyperfine sub-ensembles, which are separated by $A_{zz} = 2.1$ MHz \cite{felton2009hyperfine}.
We extract the parameters $\kappa/2 = 2\pi\times 130$ kHz, $g = 2\pi\times 190$ kHz, and $\Gamma/2 = 2\pi\times165$ kHz, where $\Gamma$ is the sub-ensemble full-width at half maximum.
Each of the three sub-ensembles reaches the strong coupling regime, $g > \kappa/2, \Gamma/2$, a first for a room-temperature sensor. 

\subsection{Nonlinear Regime}

\begin{figure}\includegraphics[width = 0.45\textwidth]{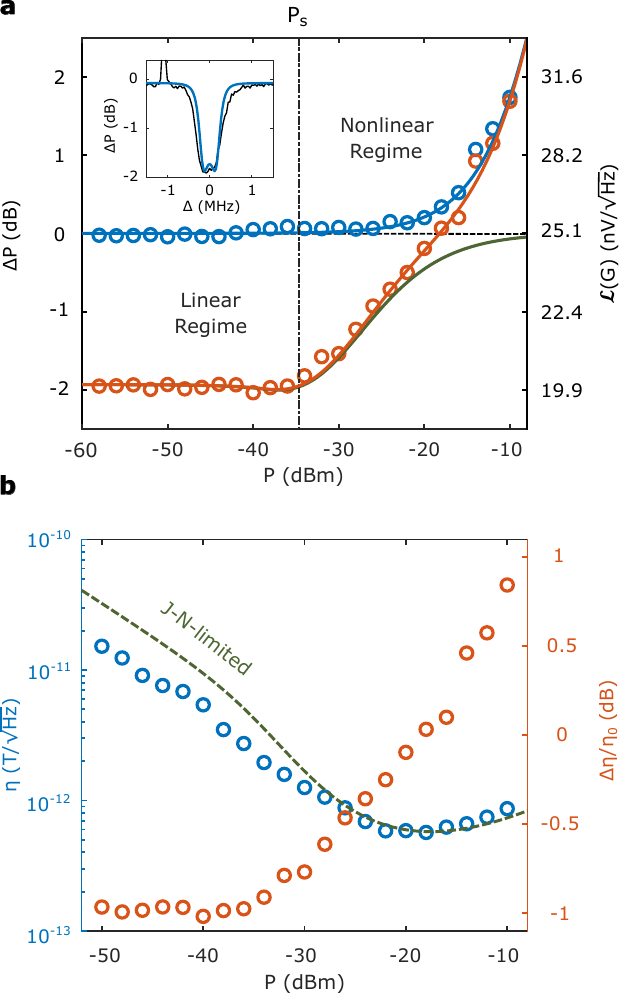}
\caption{\label{fig:3} \textbf{Spin refrigeration.} 
\textbf{a}, Noise reduction with different microwave driving powers $P$.
Blue (red) dots show the noise in 1 Hz bandwidth at 15 kHz offset in two regimes of spin-cavity detuning $\Delta_{\mathrm{s}}\neq0$ ($\Delta_{\mathrm{s}}=0$). 
The left axis shows the relative change in power spectral density compared to a 50 $\Omega$ reference measurement, and the right axis shows the voltage amplitude spectral density. All measurements were taken with a power gain of 36.5 dB and the values are doubled-sided. 
The blue curve shows the noise model considering thermal and phase noises. 
The red curve shows the noise model with linear steady-state cooling (See Methods). 
The green curve is the contribution of mode cooling alone, without thermal or phase noises.
Inset: frequency-resolved noise reduction with $P=-58$ dBm and $\Delta_{\mathrm{s}}=0$. 
\textbf{b}, Sensitivity $\eta$ at 15 kHz offset with different microwave power (blue dots).
The blue curve shows the sensitivity corresponding to the room temperature limit $\eta_0$.
The red dots show the sensitivity difference $\Delta\eta/\eta=(\eta-\eta_0)/\eta_0$.
We observe the sensor's sensitivity is sub-thermal limited with $P<-18$ dBm. }
\end{figure}

In the opposite limit of large drive power on resonance with both the cavity and the spin ensemble, the spin polarization is suppressed and the ensemble is effectively decoupled them from the resonator.
We determine the onset of this saturation behavior by examining the conditions for which $\abs{\alpha}^2\propto\abs{\beta_{\mathrm{in}}}^2$ is no longer satisfied for near resonant tuning, where the steady-state occupancy of the cavity obeys the nonlinear equation:
\begin{equation}
|\beta_\mathrm{in}|^2 = \frac{\kappa}{4}\frac{\kappa}{\kappa_{\ms{c}1}}\qty(1 
+ C_\alpha)^2\abs{\alpha}^2
\label{eq:nonlinear_cavity_number}
\end{equation}
where $C_\alpha=4g_{\mathrm{eff}}^2/\kappa\Gamma_1$ is the effective cooperativity including the nonlinear saturation effect. Here $\Gamma_1 = \Gamma +\gamma\chi$ is the effective spin linewidth and $g_{\mathrm{eff}}=g/\sqrt{\chi}$ is the effective coupling strength, modified from their linear-regime values by a factor $\chi = \sqrt{1+8g_{\mathrm{s}}^2|\alpha|^2/\gamma^2}$ that quantifies the depolarization of the spins due to the cavity field.

The cavity occupancy at which saturation begins has a natural interpretation: when the Rabi frequency of the resonant spins due to the cavity field $\sqrt{2}g_{\mathrm{s}}\abs{\alpha}$ becomes comparable to spin polarization rate $\gamma$, the system behaves nonlinearly.
For an initially polarized system, the input field at which saturation occurs is:
\begin{equation}
\abs{\beta_s}^2 = \frac{Ng^2}{2\kappa_{\ms{c}1}}\qty[\frac{\gamma}{\sqrt{2}(\Gamma + \sqrt{2}\gamma)} + \frac{\kappa\gamma}{4g^2}]^2
\label{onset}
\end{equation}
The according saturation power $P_\mathrm{s} = \hbar\omega\abs{\beta_\mathrm{s}}^2$ is plotted on Fig.~\ref{fig:2}a,c,d. Around the discussed resonant tuning point, the reflection coefficient response is confined to the quadrature channel and is given by:
\begin{align}
\pdv{\Im[r]}{\omega_\mathrm{s}} = \frac{4C_\alpha}{\Gamma_1}\frac{\kappa_{\ms{c}1}}{\kappa}\qty(1 +C_\alpha)^{-2}
\end{align}
This expression is the key result of our model; it allows us to interpret the observed optimal choice of optical polarization rate and probe power for our device, and to provide an outlook and recommendations for future device development.

\subsection{Signal Optimization}

To function as a sensor, changes in the environment that shift the NV resonance frequency must be associated with their impact on the reflected signal.
Here we measure the change in quadrature voltage as a function of the NV resonance frequency for small changes around an operating point.
We denote the constant of proportionality between the reflected voltage amplitude and the NV ensemble frequency shift as the signal $S$:
\begin{align}
S\qty(\beta_\mathrm{in}) = \abs{f(\beta_\mathrm{in})\pdv{\Im[r]}{\omega_{\mathrm{s}}}}
\label{equation:signal}
\end{align}
where $f(\beta_\mathrm{in}) = \sqrt{\hbar\omega R}\beta_\mathrm{in}$ is the input probe voltage with $R$ the resistance. 


We optimize $S$ over microwave power and optical rate for resonant tuning using the nonlinear model, as shown in Fig. \ref{fig:2}c,d,e. We find that the optimal probe power is within the nonlinear regime.
In the linear regime below saturation, the reflection coefficient is independent of the probe power and the system response, Eq.~\eqref{equation:signal}, may be increased proportional to $\abs{\beta_\mathrm{in}}$.
Conversely, for a saturated cavity and increasingly large probe strength, the reflection coefficient becomes insensitive to changes in $\omega_{\mathrm{s}}$ and the signal response decreases like $\abs{\beta_\mathrm{in}}^{-1}$ as the spin resonance is quenched and broadened.
Interestingly, if the system is sufficiently cooperative and homogeneous, $\frac{4g^2}{\kappa\gamma}\qty(\frac{\gamma}{\Gamma + \gamma})^2 \gg 1$, the saturation threshold bifurcates and the response with respect to probe power becomes hysteretic.
In this case, the optimal operating point is found on the saturated branch near the transition point, see Supplementary Material Sec.~I.

\begin{figure}
\includegraphics[width = 0.46\textwidth]{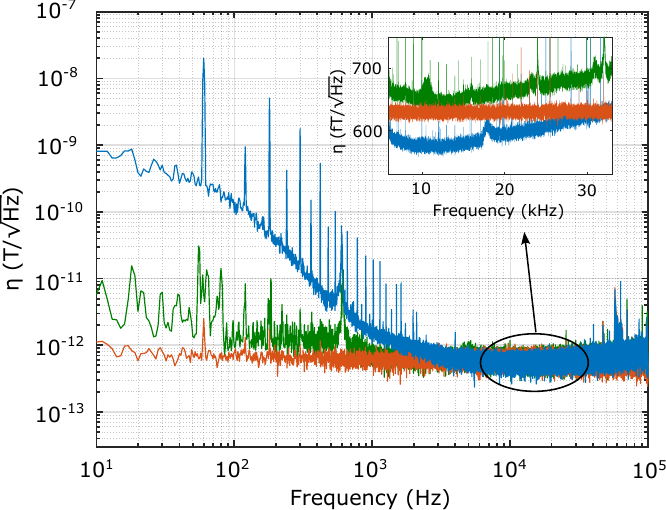}
\caption{\label{fig:4} 
\textbf{Broadband magnetometry.} 
Sensitivity of the cQED sensor as a function of magnetic field frequency derived from the voltage noise floor of the device.
Red: The magnetic-field-equivalent noise floor of the microwave circuitry, including low noise amplifier, circulator, and mixer, measured by replacing the cQED sensor with a 50 $\Omega$ terminator.
Green: The field-equivalent noise floor of the cQED sensor with NV spins detuned ($\Delta_\mathrm{s} = 5$ MHz).
Blue: Magnetic field sensitivity at the optimized operating point.
Inset: the magnetic field sensitivity around 20 kHz showing the mode cooling effect. We achieve a sub-room-temperature Johnson-Nyquist limit sensitivity of $\eta$ = 580 fT/$\sqrt{\mathrm{Hz}}$.
}
\end{figure}

The experimental response of the device is shown in Fig.~\ref{fig:2}. The reflected fraction of probe power is plotted in Fig.~\ref{fig:2}a.
For small probe strength, the probe is off-resonance with the polariton branches and is reflected.
In contrast, for large probe strength, the spin ensemble is saturated and the probe is on resonance with the now-bare cavity, which dissipates some fraction of the incident energy, suppressing the reflected signal power.
In Fig.~\ref{fig:2}B we plot experimental $\Im[r]$ for drive and spin detunings in the vicinity of the resonant operating point.
As the drive power is increased, the spin ensemble depolarizes and avoided crossing of the polariton branches are quenched.
We plot the modeled signal response of our system as a function of drive power and optical polarization rate in Fig.~\ref{fig:2}d. 
 In Fig.~\ref{fig:2}c,e we plot cuts in microwave and laser power respectively through the optimal operating point, comparing model predictions (red curve) to our experimental data.
We achieve an optimal signal response of $52~\mathrm{nV/Hz}$ at $P=-18~\mathrm{dBm}$ and $\gamma_\mathrm{p} = 2\pi \times30$~kHz, limited by the maximum output power of our polarizing laser.

In contrast to previous efforts~\cite{zhang2021exceptional, eisenach2021cavity, wilcox2022thermally}, this treatment of ensemble inhomogeneity and the optical polarization rate explains the value of the optimal microwave power and provides a better fit for our measured signal, $S$.
For comparison we plot the magnetic sensitivity predicted by a model~\cite{zhang2021exceptional} based on a Holstein-Primakoff transformation including sub-leading corrections in Fig.~\ref{fig:2}c.
Such an approach does not capture the full impact of quenching at high microwave powers and is not suited to address the role of ensemble inhomogeneity.

\section{\label{sec:3}Magnetic Sensitivity Analysis}

The sensitivity of the device may be expressed as $\eta = \sqrt{\mathcal{L}/S}/A$
for a low-frequency external stimulus and large post-reflection gain.
Here $\mathcal{L}$ is the power spectral density of voltage fluctuations exiting the cavity and, for a magnetometer, $A=\flatfrac{g\mu_\mathrm{B}}{\sqrt{3}}$ relates the external magnetic field along the diamond [100] axis to the NV-ensemble resonant frequency.
Having characterized the sensor response $S$ in the section above, we now examine the noise environment and its impact on sensitivity. 

We measure the signal, isolated in the reflected quadrature, using a saturated mixer scheme (see Methods).
Johnson-Nyquist noise typically sets the noise floor of microwave measurements. 
This voltage noise is characterized by $\mathcal{L}_\mathrm{th}= k_\mathrm{B}TR$,
where $k_\mathrm{B}$ is Boltzmann's constant, $R$ is the termination resistance, and $T\gg \omega/k_\mathrm{B}$ is the device temperature. 
In addition to the Johnson-Nyquist noise, we also experience phase noise due to fluctuations of the probe field and added noise in the homodyne measurement chain (e.g. amplifier, digitizer; see Supplementary Material Sec. II).

\begin{figure*}
\includegraphics[width = 1\textwidth]{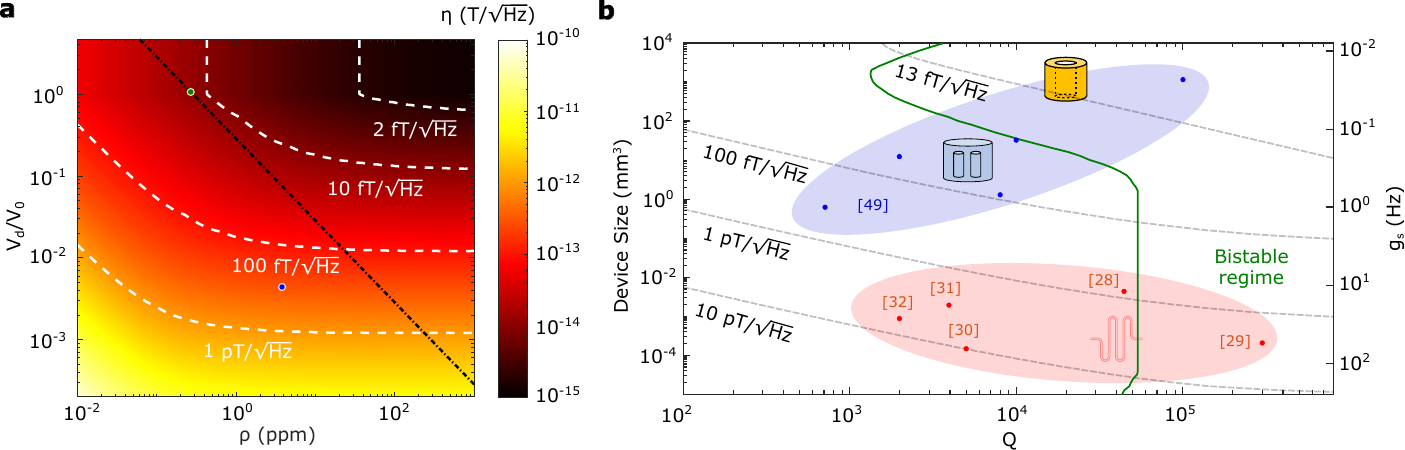}
 \caption{\label{fig:5} \textbf{Cavity and spin optimization predicted by the nonlinear model.}
 \textbf{a}, Magnetic field sensitivity with different diamond volumes $V_{\mathrm{d}}$ and NV density $\rho$. 
 The diamond volume is normalized by the cavity mode volume ($V = 1.7$ cm$^3$). 
 Four dotted lines show the contour of the sensitivity $\eta = 2~\mathrm{fT}/\sqrt{\mathrm{Hz}}$, $10~\mathrm{fT}/\sqrt{\mathrm{Hz}}$, $100~\mathrm{fT}/\sqrt{\mathrm{Hz}}$, and $1~\mathrm{pT}/\sqrt{\mathrm{Hz}}$.
 The blue dot shows the current device and the green dot shows the best sensitivity achievable from an optimized diamond using the cavity shown in this work. 
 The black dotted line shows the optical polarization limit. \textbf{b}, 
 Magnetic field sensitivity with different unloaded quality factors $Q$ and single coupling strength $g_{\mathrm{s}}$. 
 The red dots show cavity design for stripline cavities or superconducting circuits \cite{kubo2010strong,bienfait2016reaching,schuster2010high,guo2023strong,zhang2014strongly}. 
 Blue dots various three-dimensional cavity designs from \cite{choi2023ultrastrong}. The current device is shown as a yellow cylinder.}
\end{figure*}

\subsection{Spin Refrigeration}

In our system, the NV spins are polarized to the spin ground state using continuous optical pumping.
In this low entropy configuration, the spin ensemble serves as a cooling agent for the cavity mode by collectively interacting with microwave photons~\cite{fahey2023steady}, an effect seen previously in e.g. Rydberg atoms \cite{Haroche1985-in}. 
This effect is plotted in the main panel of Fig.~\ref{fig:3}a.
To establish a noise baseline, we detune the spin ensemble from the cavity by 5 MHz and measure the noise power spectral density in a 1 Hz band at 15 kHz offset in the signal channel for the bare cavity driven on resonance (blue).
At low microwave drive strengths, the microwave-power-independent thermal noise floor is characterized by $T=407$ K, including a $0.8$ dB amplifier noise figure.
Above approximately $P=-15$ dBm, the thermal noise floor is eclipsed by the phase noise of the microwave drive (See Supplementary Material Sec. II).
Bringing the spin ensemble on resonance with the cavity (Fig.~\ref{fig:3}a, red), we find a 1.98 dB suppression of the thermal noise floor at low microwave probe power, which corresponds to an effective microwave temperature reduction of $\Delta T=166$ K.
In the inset of Fig.~\ref{fig:3}a, we plot the noise power spectral density at 15 kHz offset as the microwave drive is tuned through resonance with the cavity and spin ensemble.
We find good agreement between measurements and a nonlinear cooling model (see Supplementary Material Sec. II), where the bare cavity parameters are adjusted to effective parameters using the nonlinear factor $\chi$. This spin refrigeration effect is suppressed as the same saturation onset power $P_s = -35$ dBm described in Eq. (\ref{onset}). 
At the sensitivity-maximizing operating point, $\hbar\omega\abs{\beta_\mathrm{in}}^2 = -22$ dBm, we observe a noise power reduction of 0.51 dB.

\subsection{Broadband Magnetometry}

We now infer the broadband sensitivity using the measured signal response and noise spectrum.
As shown in Fig.~\ref{fig:3}b, 
we achieve an optimal sensitivity of $\eta = 580~\mathrm{fT/\sqrt{Hz}}$ around $P=-22$ dBm and highlight it is below the ambient Johnson-Nyquist limit due to spin refrigeration.

Next we examine the spectral performance of the device under optimal probe power.
In Fig.~\ref{fig:4}, we plot the broadband magnetic sensitivity in blue.
The device achieves optimal sensitivity $\eta=$ 580 fT/$\sqrt{\mathrm{Hz}}$ between 10 -- 20 kHz.
The inset of Fig.~\ref{fig:4} shows the sensitivity near 20 kHz with an amplitude spectral density derived from the average of 3600 independently measured power spectral densities.
In the 10 -- 20 kHz band the 580 fT/$\mathrm{\sqrt{Hz}}$ device sensitivity exceeds the 620 fT/$\mathrm{\sqrt{Hz}}$ noise-equivalent sensitivity implied by the standard 50 $\Omega$ termination (red).
At frequencies below 1 kHz, ambient magnetic fields dominate the low-frequency noise and set the noise floor.
With the spin ensemble off-resonance, the sensor magnetic response is negligible and low-frequency environmental magnetic fluctuations are suppressed in the bare cavity noise-equivalent sensitivity (green).
From this spectrum, we infer a sensitivity of $\eta\approx 2~\mathrm{pT/\sqrt{Hz}}$ at around 15 Hz, neglecting possible low-frequency magnetic fluctuations intrinsic to the diamond.
To validate our noise-spectrum-inferred sensitivity, we record a noise spectrum in the presence of a test magnetic field of known amplitude.
This and further characterization of broadband sensor linearity and amplitude dynamic range is presented in the Supplementary Material Sec. IV.

\section{\label{sec:4}Sensitivity outlook}

Finally, we discuss the implications of these results for the design of future devices.
The diamond is characterized by its shape, volume $V_\mathrm{d}$, and NV density $\rho$.
The NV ensemble inhomogeneous linewidth is ultimately determined by the NV density, i.e. $\Gamma = a \rho$ with $a$ = 82.5 kHz/ppm based on our current device \cite{bauch2020decoherence}, while the homogeneous linewidth is determined by the optical pumping rate.
The resonant cavity mode is similarly characterized by its shape, volume $V$, and quality factor $Q = \omega_c/\kappa$. 
The cavity mode volume implicitly sets the single-spin coupling strength, $g_\mathrm{s}$, which, together with the ensemble size, $N=\rho V_\mathrm{d}$, determines the collective coupling strength, $g$.

We first consider the optimal diamond for our present microwave cavity mode, $V = 1.7$ cm$^3$ and $Q = 2.2\times 10^4$.
Increasing the diamond volume at fixed NV density improves device sensitivity up to unity filling factor $V_{\mathrm{d}}/V=1$, as shown in Fig.~\ref{fig:5}a, while increased density does not improve performance in the large-$\rho$ regime. 

Larger NV ensembles yield better sensitivity, but are also more challenging to optically polarize as the incident laser is attenuated within the diamond.
The optical polarization constraint $\rho_{\mathrm{max}}V_{\mathrm{d}}=0.49$ cm$^3$ ppm is satisfied to the left of the black dotted line in Fig.~\ref{fig:5}a assuming an aspect ratio of 2.2. Larger aspect ratios could relax this constraint and potentially achieve better sensitivity.
An optimal diamond (green dot) would fill the cavity mode and have the largest possible density of NV centers while achieving full optical polarization.
We predict that the resulting device should have a sensitivity of approximately 12 $\mathrm{fT/\sqrt{Hz}}$.

We next discuss alternative cavity designs with different $Q$ and $V$.
We assume that all cavities are filled with diamond at unit filling factor, $V_\mathrm{d} = V$,
and optimize the sensitivity of the device at each $Q$ and $V$ under constraints described above. The results are shown in Fig.~\ref{fig:5}b, with realistic cavity examples indicated as dots  \cite{kubo2010strong,bienfait2016reaching,schuster2010high,guo2023strong,zhang2014strongly,choi2023ultrastrong}. Two trends are clear in the results shown in Fig.~\ref{fig:5}b.
The optimal sensitivity $\eta$ is both proportional to $\sqrt{Q}$ outside of the bistable regime, and is proportional to $1/\sqrt{V}$ until the onset of optical polarization constraints.
A larger, higher quality factor device is therefore preferred for achieving optimal sensitivity.
The square-root scaling is in contrast with previous predictions that sensitivity would be proportional to device cooperativity (linear in $Q/V$).
This is because the optimal microwave power occurs at a lower value for highly cooperative devices ($P_{\mathrm{opt}} \propto \sqrt{\kappa}/g_\mathrm{s}$) and therefore limits their advantage. We discuss additional aspects of device optimization including sensitivity in the bistable regime, performance with relaxation of the optical polarization constraint, and reaching near-unity readout fidelity in the Supplementary Material Secs. I and III.

\section{Conclusion and Outlook}

The demonstrated magnetic sensitivity of 580 fT/$\sqrt{\mathrm{Hz}}$ sets the standard for the highest-performing continuous-operation NV sensor \cite{barry2020sensitivity,barry2016optical,schloss2018simultaneous}. Spin refrigeration enables high performance beyond the thermal noise that was previously thought as the fundamental limit for ambient operation, and offers a path towards quantum-limited devices at room temperature.
Our theoretical framework on the inhomogeneous nonlinear theory, validated by experiment, also allows further optimization and understanding for the operation of cQED sensors into the highly cooperative regime.
Future research will explore several avenues enabled by this proof of concept: the extension of our inhomogeneous, nonlinear theory to sensors in the bistable regime, optimization of spin refrigeration toward the quantum limit, and the exploration of other sensing modalities, such as inertial sensing and timekeeping applications.

\section{\label{sec:5}Acknowledgements\protect}
The authors would like to thank Donald Fahey, Reginald Wilcox, David Phillips, Danielle Braje, Andrew Kerman, and Avetik Harutyunyan for helpful discussions.
H.W.~acknowledges support from Analog Devices, Inc. and Honda Research Institute USA, Inc. D.R.E.~acknowledges funding from the MITRE Corporation and the U.S.~NSF Ce
nter for Ultracold Atoms. 

This material is based upon work supported by the Dept.~of the Army under Air Force Contract No.~FA8702-15-D-0001.
Any opinions, findings, conclusions or recommendations expressed in this material are those of the author(s) and do not necessarily reflect the views of the Dept.~of the Army.

\section{Data availability}
The data that supports the findings of this study is available from the corresponding author upon reasonable request.

\section{Code availability}
The code that supports the findings of this study is available from the corresponding author upon reasonable request.

\section{Competing interests}
The authors declare that they have no competing interests.

\section{Author Contribution}
H.W. and M.E.T. created the setup and conducted the experiments. D.R.E. proposed the spin refrigeration for the cQED sensor.
H.W., K.L.T., K.J., D.R.E, and M.E.T. developed the nonlinear model.
H.W., K.L.T, and M.E.T. prepared the manuscript. X.Z. and M.J. assisted in electrical measurement design.
All authors discussed the results and revised the manuscript.
M.E.T. and D.R.E. supervised the project.

\section{Methods}

\subsection{Experimental setup}

We employ a saturated mixer scheme to measure the phase change induced by the NV-cavity system. 
Probe microwaves from a signal generator are divided into a reference arm and a signal arm using a Wilkinson microwave power divider.
The reference arm is directly connected to the LO port of a mixer (HX3400), with a voltage-controlled phase shifter to tune the relative phase.
The signal arm is directed to a circulator with tunable attenuation for power control on the cavity input, and the circulator's microwave output is coupled into the dielectric resonator using a probe loop. The reflected signal from the cavity returns to the circulator and is connected to a low-noise amplifier and subsequently the RF port of the mixer for a homodyne measurement.
This setup effectively separates the reflected microwave signal from the incident signal, enabling the measurement of the quadrature part of the reflection coefficient by appropriate setting of the LO phase shifter. The quadrature signal is subsequently digitized using a sampling rate of 200 kS/s. Power spectral density in Fig. 3 and Fig. 4 is calculated with Welch’s power spectral density estimate. For power measurements (Fig 1c), an IQ mixer is used to measure both field quadratures, and the total power is then computed.

The diamonds (3 mm $\times$ 3 mm $\times$ 0.9 mm, 4 ppm NV ensemble, sourced from Element 6) are set at the TE01$\delta$ mode maximum point in the center of the dielectric resonator (Skyworks, $\varepsilon_\mathrm{r} \sim 31$).
A wafer of 4H-SiC is used for heat transfer and supporting the diamond, while two pieces of low-loss-tangent polytetrafluoroethylene (PTFE) are used to fix and align the dielectric resonator.
An aluminum shield is employed to isolate the system from external signals in the lab, such as WiFi and 3G signals (1.9 GHz), and to reduce radiative losses.
An 8W 532 nm pump laser is utilized to optically polarize the spin ensemble. External magnetic bias is provided by 3-axis magnetic coils.

\subsection{Test field}
We performed a test magnetic field measurement using a coil with turn number $N= 400$, radius $r=7$ cm, and distance between the coil and sensor $d=22$ cm. We use a DC power supply to generate a voltage $U=1$ V, resulting in a current output of $I=43$ mA. The magnetic field generated by this test coil at the sensor position is $B_{\mathrm{cal}}=2\mu_0\pi r^2NI/4\pi(d^2+r^2)^{3/2} = 4.3~\mu T$, which is similar with the measurement from the measurement from a commercial gaussmeter $B_{\mathrm{Gauss}} = 3.9~\mu T$. Our sensor shows a response of $B_{\mathrm{cQED}} = 4.0~\mu T$ with a 2.5\% (7\%) error compared with the reference magnetometer (calculated result). The difference can be attributed to the estimation of the size parameters and the twisted angle of the coil with the diamond surface. 

\subsection{Sensitivity prediction}
We applied the first-order approximation (See Supplementary Materials Sec. I) for the sensitivity optimization for Fig. 5. We then collected the optimal parameters and determined the number of solutions. Following this, we identified and marked the bistability regime in Fig. 5b, which is characterized by its multiple solutions. The parameters for optimization and the constraints are listed and plotted in Sec. IV of the Supplementary Materials.

\bibliography{apssamp}

\end{document}